\documentclass[runningheads]{llncs}

\usepackage{graphicx}
\usepackage{hyperref}
\usepackage{xcolor}
\usepackage{textcomp}
\usepackage{amsfonts}
\usepackage{amsmath}
\usepackage{amssymb}
\usepackage{mathtools}
\usepackage{mwe}
\usepackage{cite}

\hypersetup{
	colorlinks,
	linkcolor={red!50!black},
	citecolor={blue!60!black},
	urlcolor={blue!80!black}
}

\begin{document}

\title{Discovering Process Models from\\ Uncertain Event Data\thanks{In \emph{International Workshop on Business Process Intelligence} (BPI 2019). DOI: 10.1007/978-3-030-37453-2\_20. \textcopyright Springer. We thank the Alexander von Humboldt (AvH) Stiftung for supporting our research interactions. Please do not print this document unless strictly necessary.}}

\author{Marco Pegoraro\orcidID{0000-0002-8997-7517} \and Merih Seran Uysal\orcidID{0000-0003-1115-6601} \and Wil M.P. van der Aalst\orcidID{0000-0002-0955-6940}}

\authorrunning{Pegoraro et al.}

\institute{Process and Data Science Group (PADS) \\ Department of Computer Science, RWTH Aachen University, Aachen, Germany
	\email{\{pegoraro,uysal,wvdaalst\}@pads.rwth-aachen.de}\\
	\url{http://www.pads.rwth-aachen.de/}}

\maketitle

\begin{abstract}
	Modern information systems are able to collect event data in the form of \emph{event logs}. Process mining techniques allow to discover a model from event data, to check the conformance of an event log against a reference model, and to perform further process-centric analyses. In this paper, we consider uncertain event logs, where data is recorded together with explicit uncertainty information. We describe a technique to discover a directly-follows graph from such event data which retains information about the uncertainty in the process. We then present expe\-rimental results of performing inductive mining over the directly-follows graph to obtain models representing the certain and uncertain part of the process.
	
\end{abstract}

\section{Introduction}
With the advent of digitalization of business processes and related management tools, \emph{Process-Aware Information Systems} (PAISs), ranging from ERP/CRM-systems to BPM/WFM-systems, are widely used to support operational admi\-nistration of processes. The databases of PAISs containing event data can be queried to obtain \emph{event logs}, collections of recordings of the execution of activities belonging to the process. The discipline of \emph{process mining} aims to synthesize knowledge about \emph{processes} via the extraction and analysis of execution logs.

When applying process mining in real-life settings, the need to address anomalies in data recording when performing analyses is omnipresent. A number of such anomalies can be modeled by using the notion of uncertainty: \emph{uncertain event logs} contain, alongside the event data, some attributes that describe a certain level of uncertainty affecting the data. A typical example is the timestamp information: in many processes, specifically the ones where data is in part manually recorded, the timestamp of events is recorded with low precision (e.g., specifying only the day of occurrence). If multiple events belonging to the same case are recorded within the same time unit, the information regarding the event order is lost. This can be modeled as uncertainty of the timestamp attribute by assigning a time interval to the events. Another example of uncertainty are situations where the activity label is unrecorded or lost, but the events are associated with specific resources that carried out the corresponding activity. In many organizations, each resource is authorized to perform a limited set of activities, depending on her role. In this case, it is possible to model the absence of activity labels associating every event with the set of possible activities which the resource is authorized to perform.

Usually, information about uncertainty is not natively contained into a log: event data is extracted from information systems as activity label, timestamp and case id (and possibly additional attributes), without any sort of meta-information regarding uncertainty. In some cases, a description of the uncertainty in the process can be obtained from background knowledge. Information translatable to uncertainty such as the one given above as example can, for instance, be acquired from an interview with the process owner, and then inserted in the event log with a pre-processing step. Research efforts regarding how to discover uncertainty in a representation of domain knowledge and how to translate it to obtain an uncertain event log are currently ongoing.

Uncertainty can be addressed by filtering out the affected events when it appears sporadically throughout an event log. Conversely, in situations where uncertainty affects a significant fraction of an event log, filtering out uncertain events can lead to information loss such that analysis becomes very difficult. In this circumstance, it is important to deploy process mining techniques that allow to mine information also from the uncertain part of the process.

In this paper, we aim to develop a process discovery approach for uncertain event data. We present a methodology to obtain \emph{Uncertain Directly-Follows Graphs} (UDFGs), models based on directed graphs that synthesize information about the uncertainty contained in the process. We then show how to convert UDFGs in models with execution semantics via filtering on uncertainty information and inductive mining.

The remainder of the paper is structured as follows: in Section~\ref{sec:related} we present relevant previous work. In Section~\ref{sec:preliminaries}, we provide the preliminary information necessary for formulating uncertainty. In Section~\ref{sec:udfg}, we define the uncertain version of directly-follows graphs. In Section~\ref{sec:uimd}, we describe some examples of exploiting UDFGs to obtain executable models. Section~\ref{sec:exp} presents some experiments. Section~\ref{sec:conc} proposes future work and concludes the paper.

\section{Related Work}\label{sec:related}
In a previous work~\cite{pegoraro2019mining}, we proposed a taxonomy of possible types of uncertainty in event data. To the best of our knowledge, no previous work addressing explicit uncertainty currently exist in process mining. Since usual event logs do not contain any hint regarding misrecordings of data or other anomalies, the notion of ``noise'' or ``anomaly'' normally considered in process discovery refers to outlier behavior. This is often obtained by setting thresholds to filter out the behavior not considered for representation in the resulting process model. A variant of the Inductive Miner by Leemans et al.~\cite{leemans2013discovering} considers only directly-follows relationships appearing with a certain frequency. In general, a direct way to address infrequent behavior on the event level is to apply on it the concepts of support and confidence, widely used in association rule learning~\cite{hornik2005arules}. More sophisticated techniques employ infrequent pattern detection employing a mapping between events~\cite{lu2016detecting} or a finite state automaton~\cite{conforti2017filtering} mined from the most frequent behavior.

Although various interpretations of uncertain information can exist, this paper presents a novel approach that aims to represent uncertainty explicitly, rather than filtering it out. For this reason, existing approaches to identify noise cannot be applied to the problem at hand. 

\section{Preliminaries}\label{sec:preliminaries}
To define uncertain event data, we introduce some basic notations and concepts, partially from~\cite{van2016process}:

\begin{definition}[Power Set]
	The power set of a set $A$ is the set of all possible subsets of $A$, and is denoted with $\mathcal{P}(A)$. $\mathcal{P}_{NE}(A)$ denotes the set of all the non-empty subsets of $A$: $\mathcal{P}_{NE}(A) = \mathcal{P}(A)\setminus\{\emptyset\}$.
\end{definition}

\begin{definition}[Sequence]
	Given a set $X$, a finite \emph{sequence} over $X$ of length $n$ is a function $s \in X^* : \{1, \dots, n\} \rightarrow X$, typically written as $s = \langle s_1, s_2, \dots, s_n\rangle$. For any sequence $s$ we define $|s| = n$, $s[i] = s_i$, $\mathcal{S}_s = \{s_1, s_2, \dots, s_n\}$ and $x \in s \iff x \in \mathcal{S}_s$. Over the sequences $s$ and $s'$ we define $s \cup s' = \{a\in s\} \cup \{a\in s'\}$.
\end{definition}

\begin{definition}[Directed Graph]
	A \emph{directed graph} $G = (V, E)$ is a set of \emph{vertices} $V$ and a set of directed \emph{edges} $E \subseteq V \times V$. We denote with $\mathcal{U}_G$ the universe of such directed graphs.
\end{definition}

\begin{definition}[Bridge]
	An edge $e \in E$ is called a \emph{bridge} if and only if the graph becomes disconnected if $e$ is removed: there exists a partition of $V$ into $V'$ and $V''$ such that $E \cap ((V' \times V'') \cup (V''\times V')) = \{ e \}$. We denote with $E_B \subseteq E$ the set of all such bridges over the graph $G = (V, E)$.
\end{definition}

\begin{definition}[Path]
	A \emph{path} over a graph $G = (V, E)$ is a sequence of vertices $p = \langle v_1, v_2, \dots v_n \rangle$ with $v_1, \dots, v_n \in V$ and $\forall_{1 \leq i \leq n-1} (v_i, v_{i+1}) \in E$. $P_G(v, w)$ denotes the set of all paths connecting $v$ and $w$ in $G$. A vertex $w \in V$ is \emph{reachable} from $v \in V$ if there is at least one path connecting them: $|P_G(v, w)|>0$.
\end{definition}

\begin{definition}[Transitive Reduction]
	A \emph{transitive reduction} of a graph $G = (V, E)$ is a graph $\rho(G) = (V, E')$ with the same reachability between vertices and a minimal number of edges. $E' \subseteq E$ is a smallest set of edges such that $|P_{\rho(G)}(v, w)|>0 \implies |P_G(v, w)|>0$ for any $v, w \in V$.
\end{definition}

In this paper, we consider \emph{uncertain event logs}. These event logs contain uncertainty information explicitly associated with event data. A taxonomy of different kinds of uncertainty and uncertain event logs has been presented in~\cite{pegoraro2019mining} which it distinguishes between two main classes of uncertainty. \emph{Weak uncertainty} provides a probability distribution over a set of possible values, while \emph{strong uncertainty} only provides the possible values for the corresponding attribute.

We will use the notion of \emph{simple uncertainty}, which includes strong uncertainty on the control-flow perspective: activities, timestamps, and indeterminate events. An example of a simple uncertain trace is shown in Table~\ref{table:uncertaintrace}. Event $e_1$ has been recorded with two possible activity labels ($a$ or $c$), an example of strong uncertainty on activities. Some events, e.g. $e_2$, do not have a precise timestamp but a time interval in which the event could have happened has been recorded: in some cases, this causes the loss of the precise order of events (e.g. $e_1$ and $e_2$). These are examples of strong uncertainty on timestamps. As shown by the ``?'' symbol, $e_3$ is an indeterminate event: it has been recorded, but it is not guaranteed to have happened.

\begin{table}[]
	\caption{An example of simple uncertain trace.}
	\label{table:uncertaintrace}
	\centering
	\begin{tabular}{ccccc}
		\textbf{Case ID} & \textbf{Event ID}        & \textbf{Activity}                                                                                                     & \textbf{Timestamp}             & \multicolumn{1}{l}{\textbf{Event Type}} \\ \hline
		\multicolumn{1}{|c|}{0} & \multicolumn{1}{|c|}{$e_1$} &
		\multicolumn{1}{c|}{\{a, c\}} & \multicolumn{1}{c|}{\begin{tabular}[c]{@{}c@{}}[2011-12-02T00:00\\ 2011-12-05T00:00]\end{tabular}}                                                                                 & \multicolumn{1}{c|}{!}                    \\ \hline
		\multicolumn{1}{|c|}{0} & \multicolumn{1}{|c|}{$e_2$} &
		\multicolumn{1}{c|}{\{a, d\}} & \multicolumn{1}{c|}{\begin{tabular}[c]{@{}c@{}}[2011-12-03T00:00\\ 2011-12-05T00:00]\end{tabular}}                                                                         &  \multicolumn{1}{c|}{!}                    \\ \hline
		\multicolumn{1}{|c|}{0} & \multicolumn{1}{|c|}{$e_3$} &
		\multicolumn{1}{c|}{\{a, b\}}        & \multicolumn{1}{c|}{2011-12-07T00:00} &  \multicolumn{1}{c|}{?}                    \\ \hline
		\multicolumn{1}{|c|}{0} & \multicolumn{1}{|c|}{$e_4$} &
		\multicolumn{1}{c|}{\{a, b\}} & \multicolumn{1}{c|}{\begin{tabular}[c]{@{}c@{}}[2011-12-09T00:00\\ 2011-12-15T00:00]\end{tabular}}                                                                         &  \multicolumn{1}{c|}{!}                    \\ \hline
		\multicolumn{1}{|c|}{0} & \multicolumn{1}{|c|}{$e_5$} &
		\multicolumn{1}{c|}{\{b, c\}}        & \multicolumn{1}{c|}{\begin{tabular}[c]{@{}c@{}}[2011-12-11T00:00\\ 2011-12-17T00:00]\end{tabular}}                                                                         &  \multicolumn{1}{c|}{!}                    \\ \hline
		\multicolumn{1}{|c|}{0} & \multicolumn{1}{|c|}{$e_6$} &
		\multicolumn{1}{c|}{\{b\}}        & \multicolumn{1}{c|}{2011-12-20T00:00}                                                                         &  \multicolumn{1}{c|}{!}                    \\ \hline
	\end{tabular}
\end{table}

\begin{definition}[Universes]
	Let $\mathcal{U}_E$ be the set of all the \emph{event identifiers}. Let $\mathcal{U}_C$ be the set of all \emph{case ID identifiers}. Let $\mathcal{U}_A$ be the set of all the \emph{activity identifiers}. Let $\mathcal{U}_T$ be the totally ordered set of all the \emph{timestamp identifiers}. Let $\mathcal{U}_O = \{!, ?\}$, where the ``!'' symbol denotes \emph{determinate events}, and the ``?'' symbol denotes \emph{indeterminate events}.
\end{definition}

\begin{definition}[Simple uncertain traces and logs]
	$\sigma \in \mathcal{P}_{NE}(\mathcal{U}_E \times \mathcal{P}_{NE}(\mathcal{U}_A) \times \mathcal{U}_T \times \mathcal{U}_T \times \mathcal{U}_O)$ is a \emph{simple uncertain trace} if for any $(e_i, A, t_{min}, t_{max}, o) \in \sigma$, $t_{min} < t_{max}$ and all the event identifiers are unique. $\mathcal{T}_U$ denotes the universe of simple uncertain traces. $L \in \mathcal{P}(\mathcal{T}_U)$ is a \emph{simple uncertain log} if all the event identifiers in the log are unique. Over the uncertain event $e = (e_i, A, t_{min}, t_{max}, o) \in \sigma$ we define the following projection functions: $\pi_A(e) = A$, $\pi_{t_{min}}(e) = t_{min}$, $\pi_{t_{max}}(e) = t_{max}$ and $\pi_o(e) = o$. Over $L \in \mathcal{P}(\mathcal{T}_U)$ we define the following projection function: $\Pi_A(L) = \bigcup_{\sigma \in L}\bigcup_{e \in \sigma}\pi_A(e)$.
\end{definition}

The behavior graph is a structure that summarizes information regarding the uncertainty contained in a trace. Namely, two vertices are linked by an edge if their corresponding events may have happened one immediately after the other.

\begin{definition}[Behavior Graph]
	Let $\sigma \in \mathcal{T}_U$ be a simple uncertain trace. A \emph{behavior graph} $\beta \colon \mathcal{T}_U \to \mathcal{U}_G$ is the transitive reduction of a directed graph $\rho(G)$, where $G = (V, E) \in \mathcal{U}_G$ is defined as:
	\begin{itemize}
		\item $V = \{e \in \sigma \}$
		\item $E = \{(v, w) \mid v, w \in V \wedge \pi_{t_{max}}(v) < \pi_{t_{min}}(w)\}$
	\end{itemize}
\end{definition}

\begin{figure}
	\centering
	\includegraphics[width=0.5\textwidth, keepaspectratio, angle=-90]{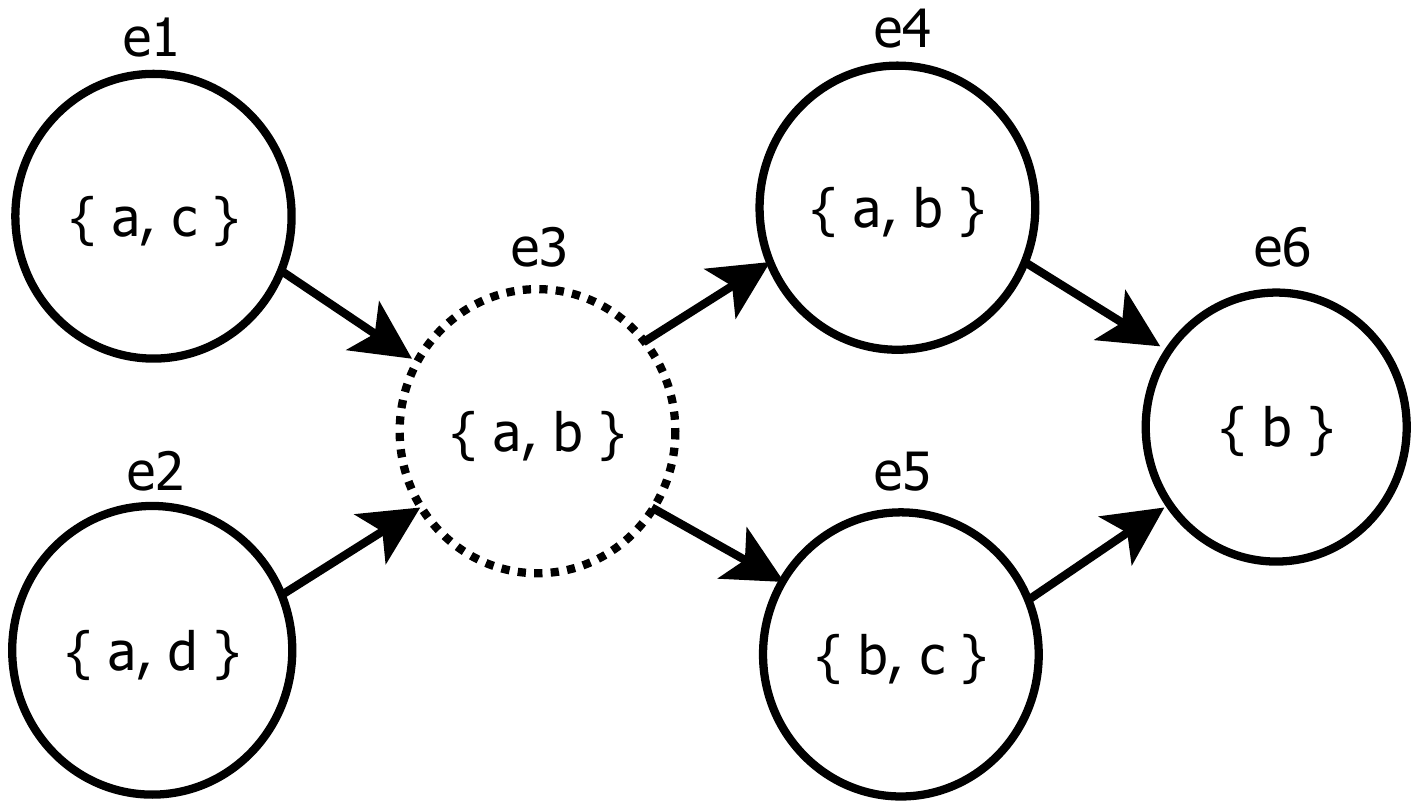}
	\caption{The behavior graph of the uncertain trace given in Table~\ref{table:uncertaintrace}. Each vertex represents an uncertain event and is labeled with the possible activity label of the event. The dotted circle represents an indeterminate event (may or may not have happened).}
	\label{fig:graphred}
\end{figure}

Notice that the behavior graph is obtained from the transitive reduction of an acyclic graph, and thus is unique. The behavior graph for the trace in Table~\ref{table:uncertaintrace} is shown in Figure~\ref{fig:graphred}.

\section{Uncertain DFGs}\label{sec:udfg}
The definitions shown in Section~\ref{sec:preliminaries} allow us to introduce some fundamental concepts necessary to perform discovery in an uncertain setting. Let us define a measure for the frequencies of single activities. In an event log without uncertainty the frequency of an activity is the number of events that have the corresponding activity label. In the uncertain case, there are events that can have multiple possible activity labels. For a certain activity $a \in \mathcal{U}_A$, the minimum activity frequency of $a$ is the number of events that certainly have $A$ as activity label and certainly happened; the maximum activity frequency is the number of events that may have $A$ as activity label.

\begin{definition}[Minimum and maximum activity frequency]
	The \emph{minimum} and \emph{maximum activity frequency} $\#_\text{min} \colon \mathcal{T}_U \times {\mathcal{U}_A} \to \mathbb{N}$ and $\#_\text{max} \colon \mathcal{T}_U \times {\mathcal{U}_A} \to \mathbb{N}$ of an activity $a \in \mathcal{U}_A$ in regard of an uncertain trace $\sigma \in \mathcal{T}_U$ are defined as:
	\begin{itemize}
		\item $\#_\text{min}(\sigma, a) = |\{e \in \sigma \mid \pi_A(e) = \{a\} \wedge \pi_o(v) = \:! \}|$
		\item $\#_\text{max}(\sigma, a) = |\{e \in \sigma \mid a \in \pi_A(e) \}|$.
	\end{itemize}
\end{definition}

Many discovery algorithms exploit the concept of \emph{directly-follows relationship}~\cite{van2004workflow, leemans2013discovering}. In this paper, we extend this notion to uncertain traces and uncertain event logs. An uncertain trace embeds some behavior which depends on the ins\-tantiation of the stochastic variables contained in the event attributes. Some directly-follows relationships exist in part, but not all, the possible behavior of an uncertain trace. As an example, consider events $e_3$ and $e_5$ in the uncertain trace shown in Table~\ref{table:uncertaintrace}: the relationship ``$a$ is directly followed by $b$'' appears once only if $e_3$ actually happened immediately before $e_5$ (i.e., $e_4$ did not happen in-between), and if the activity label of $e_3$ is a $b$ (as opposed to $c$, the other possible label). In all the behavior that does not satisfy these conditions, the directly-follows relation does not appear on $e_3$ and $e_5$.

Let us define as \emph{realizations} all the possible certain traces that are obtainable by choosing a value among all possible ones for an uncertain attribute of the uncertain trace. For example, some possible realizations of the trace in Table~\ref{table:uncertaintrace} are $\langle a, d, b, a, c, b \rangle$, $\langle a, a, a, a, b, b \rangle$, and $\langle c, a, c, b, b \rangle$. We can express the strength of the directly-follows relationship between two activities in an uncertain trace by counting the minimum and maximum number of times the relationship can appear in one of the possible realizations of that trace. To this goal, we exploit some structural properties of the behavior graph in order to obtain the minimum and maximum frequency of directly-follows relationships in a simpler manner.

A useful property to compute the minimum number of occurrences between two activities exploits the fact that parallel behavior is represented by the branching of arcs in the graph. Two connected determinate events have happened one immediately after the other if the graph does not have any other parallel path: if two determinate events are connected by a bridge, they will certainly happen in succession. This property is used to define a \emph{strong sequential relationship}.

The next property accounts for the fact that, by construction, uncertain events corresponding to nodes in the graph not connected by a path can happen in any order. This follows directly from the definition of the edges in the graph, together with the transitivity of $\mathcal{U}_T$ (which is a totally ordered set). This means that two disconnected nodes $v$ and $w$ may account for one occurrence of the relation ``$\pi_A(v)$ is directly followed by $\pi_A(w)$''. Conversely, if $w$ is reachable from $v$, the directly-follows relationship may be observed if all the events separating $v$ from $w$ are indeterminate (i.e., there is a chance that no event will interpose between the ones in $v$ and $w$). This happens for vertices $e_2$ and $e_4$ in the graph in Figure~\ref{fig:graphred}, which are connected by a path and separated only by vertex $e_3$, which is indeterminate. This property is useful to compute the maximum number of directly-follows relationships between two activities, leading to the notion of \emph{weak sequential relationship}.

\begin{definition}[Strong sequential relationship]
	Given a behavior graph $\beta = (V, E)$ and two vertices $v, w \in V$, $v$ is in a \emph{strong sequential relationship} with $w$ (denoted by $v \blacktriangleright_{\beta} w$) if and only if $\pi_o(v) = \:!$ and $\pi_o(w) = \:!$ ($v$ and $w$ are both determinate) and there is a bridge between them: $(v, w) \in E_B$.
\end{definition}

\begin{definition}[Weak sequential relationship]
	Given a behavior graph $\beta = (V, E)$ and two vertices $v, w \in V$, $v$ is on a \emph{weak sequential relationship} with $w$ (denoted by $v \triangleright_{\beta} w$) if and only if $|P_{\beta}(w, v)| = 0$ ($v$ is unreachable from $w$) and no node in any possible path between $v$ and $w$, excluding $v$ and $w$, is determinate: $\bigcup_{p \in P_\beta(v,w)} \{e \in p \mid \pi_o(e) = \:! \} \setminus \{v, w\} = \emptyset$.
\end{definition}

Notice that if $v$ and $w$ are mutually unreachable they are also in a mutual weak sequential relationship. Given two activity labels, these properties allow us to extract sets of candidate pairs of vertices of the behavior graph.

\begin{definition}[Candidates for minimum and maximum directly-follows frequencies]
	Given two activities $a, b \in \mathcal{U}_A$ and an uncertain trace $\sigma \in \mathcal{T}_U$ and the corresponding behavior graph $\beta(\sigma) = (V, E)$, the \emph{candidates for minimum and maximum directly-follows frequency} $\text{cand}_\text{min} \colon \mathcal{T}_U \times \mathcal{U}_A \times \mathcal{U}_A \to \mathcal{P}(V \times V)$ and $\text{cand}_\text{max} \colon \mathcal{T}_U \times \mathcal{U}_A \times \mathcal{U}_A \to \mathcal{P}(V \times V)$ are defined as:
	\begin{itemize}
		\item $\text{cand}_\text{min}(\sigma, a, b) = \{(v, w) \in V \times V \mid v \neq w \wedge \pi_A(v) = \{a\} \wedge \pi_A(w) = \{b\} \wedge v \blacktriangleright_{\beta} w \}$
		\item $\text{cand}_\text{max}(\sigma, a, b) = \{(v, w) \in V \times V \mid v \neq w \wedge a \in \pi_A(v) \wedge b \in \pi_A(w) \wedge v	\triangleright_{\beta} w \}$
	\end{itemize}
\end{definition}

After obtaining the sets of candidates, it is necessary to select a subset of pair of vertices such that there are no repetitions. In a realization of an uncertain trace, an event $e$ can only have one successor: if multiple vertices of the behavior graph correspond to events that can succeed $e$, only one can be selected.

Consider the behavior graph in Figure~\ref{fig:graphred}. If we search candidates for ``$a$ is directly followed by $b$'', we find $\text{cand}_\text{min}(\sigma, a, b) = \{(e_1,e_3), (e_2,e_3), (e_1,e_5), \allowbreak (e_2,e_4), (e_3,e_4), (e_3,e_5), (e_4,e_6)\}$. However, there are no realizations of the trace represented by the behavior graph that contains all the candidates; this is because some vertices appear in multiple candidates. A possible realization with the highest frequency of $a \rightarrow b$ is $\langle d, a, b, c, a, b \rangle$. Conversely, consider ``$a$ is directly followed by $a$''. When the same activity appears in both sides of the relationship, an event can be part of two different occurrences, as first member and second member; e. g., in the trace $\langle a, a, a \rangle$, the relationship $a \rightarrow a$ occurs two times, and the second event is part of both occurrences. In the behavior graph of Figure~\ref{fig:graphred}, the relation $a \rightarrow b$ cannot be supported by candidates $(e_1, e_3)$ and $(e_3, e_4)$ at the same time, because $e_3$ has either label $a$ or $b$ in a realization. But $(e_1, e_3)$ and $(e_3, e_4)$ can both support the relationship $a \rightarrow a$, in realizations where $e_1$, $e_3$ and $e_4$ all have label $a$.

When counting the frequencies of directly follows relationships between the activities $a$ and $b$, every node of the behavior graph can appear at most once if $a \neq b$. If $a = b$, every node can appear once on each side of the relationship.

\begin{definition}[Minimum directly-follows frequency]
	Given $a, b \in \mathcal{U}_A$ and $\sigma \in \mathcal{T}_U$, let $R_\text{min} \subseteq \text{cand}_\text{min}(\sigma, a, b)$ be a largest set such that for any $(v, w), (v', w') \in R_\text{min}$, it holds:
	\[
	(v, w) \neq (v', w') \implies \{v,w\} \cap \{v',w'\} = \emptyset,	\qquad	\text{if} \ a \neq b
	\]
	\[
	(v, w) \neq (v', w') \implies v \neq v' \wedge w \neq w',	\qquad	\text{if} \ a = b
	\]
	The \emph{minimum directly-follows frequency} $\rightsquigarrow_\text{min} \colon \mathcal{T}_U \times {\mathcal{U}_A}^2 \to \mathbb{N}$ of two activities $a, b \in \mathcal{U}_A$ in regard of an uncertain trace $\sigma \in \mathcal{T}_U$ is defined as $\rightsquigarrow_\text{min}(\sigma, a, b) = |R_\text{min}|$.
\end{definition}

\begin{definition}[Maximum directly-follows frequency]
	Given $a, b \in \mathcal{U}_A$ and $\sigma \in \mathcal{T}_U$, let $R_\text{max} \subseteq \text{cand}_\text{max}(\sigma, a, b)$ be a largest set such that for any $(v, w), (v', w') \in R_\text{max}$, it holds:
	\[
	(v, w) \neq (v', w') \implies \{v,w\} \cap \{v',w'\} = \emptyset,	\qquad	\text{if} \ a \neq b
	\]
	\[
	(v, w) \neq (v', w') \implies v \neq v' \wedge w \neq w',	\qquad	\text{if} \ a = b
	\]
	The \emph{maximum directly-follows frequency} $\rightsquigarrow_\text{max} \colon \mathcal{T}_U \times {\mathcal{U}_A}^2 \to \mathbb{N}$ of two activities $a, b \in \mathcal{U}_A$ in regard of an uncertain trace $\sigma \in \mathcal{T}_U$ is defined as $\rightsquigarrow_\text{max}(\sigma, a, b) = |R_\text{max}|$.
\end{definition}

For the uncertain trace in Table~\ref{table:uncertaintrace}, $\rightsquigarrow_\text{min}(\sigma, a, b) = 0$, because $R_\text{min} = \emptyset$; conversely, $\rightsquigarrow_\text{max}(\sigma, a, b) = 2$, because a maximal set of candidates is $R_\text{max} = \{(e_1, e_3), (e_4, e_6)\}$. Notice that maximal candidate sets are not necessarily unique: $R_\text{max} = \{(e_2, e_3), (e_4, e_6)\}$ is also a valid one.

The operator $\rightsquigarrow$ synthesizes information regarding the strength of the directly-follows relation between two activities in an event log where some events are uncertain. The relative difference between the \emph{min} and \emph{max} counts is a measure of how certain the relationship is when it appears in the event log. Notice that, in the case where no uncertainty is contained in the event log, \emph{min} and \emph{max} will coincide, and will both contain a directly-follows count for two activities.

An \emph{Uncertain DFG} (UDFG) is a graph representation of the activity frequencies and the directly-follows frequencies; using the measures we defined, we exclude the activities and the directly-follows relations that never happened.

\begin{definition}[Uncertain Directly-Follows Graph (UDFG)]
	Given an event log $L \in \mathcal{P}(\mathcal{T}_U)$, the Uncertain Directly-Follows Graph $DFG_{\text{U}}(L)$ is a directed graph $G = (V, E)$ where:
	\begin{itemize}
		\item $V = \{a \in \Pi_A(L) \mid \sum_{\sigma \in L} \#_\text{max}(\sigma, a) > 0 \}$
		\item $E = \{(a, b) \in V \times V \mid \sum_{\sigma \in L} \rightsquigarrow_\text{max}(\sigma, a, b) > 0 \}$
	\end{itemize}
\end{definition}

The UDFG is a low-abstraction model that, together with the data decorating vertices and arcs, gives indications on the overall uncertainty affecting activities and directly-follows relationships. Moreover, the UDFG does not filter out uncertainty: the information about the uncertain portion of a process is summarized by the data labeling vertices and edges. In addition to the elimination of the anomalies in an event log in order to identify the happy path of a process, this allows the process miner to isolate the uncertain part of a process, in order to study its features and analyze its causes. In essence however, this model has the same weak points as the classic DFG: it does not support concurrency, and if many activities happen in different order the DFG creates numerous loops that cause underfitting.

\section{Inductive Mining Using Directly-Follows Frequencies}\label{sec:uimd}
A popular process mining algorithm for discovering executable models from DFGs is the Inductive Miner~\cite{leemans2013discovering}. A variant presented by Leemans et al.~\cite{leemans2018scalable}, the \emph{Inductive Miner--directly-follows} ($\text{IM}_{D}$), has the peculiar feature of preprocessing an event log to obtain a DFG, and then discover a process tree exclusively from the graph, which can then be converted to a Petri net. This implies a high scalability of the algorithm, which has a linear computational cost over the number of events in the log, but it also makes it suited to the case at hand in this paper. To allow for inductive mining, and subsequent representation of the process as a Petri net, we introduce a form of filtering called UDFG slicing, based on four filtering parameters: $act_{\textit{min}}$, $act_{\textit{max}}$, $rel_{\textit{min}}$ and $rel_{\textit{max}}$. The parameters $act_{\textit{min}}$ and $act_{\textit{max}}$ allow to filter on nodes of the UDFG, based on how certain the corresponding activity is in the log. Conversely, $rel_{\textit{min}}$ and $rel_{\textit{max}}$ allow to filter on edges of the UDFG, based on how certain the corresponding directly-follows relationship is in the log.

\begin{definition}[Uncertain DFG slice]
	Given an uncertain event log $L \in \mathcal{P}(\mathcal{T}_U)$, its uncertain directly-follows graph $DFG_{\text{U}}(L) = (V', E')$, and $act_{\text{min}}, \allowbreak act_{\text{max}}, rel_{\text{min}}, rel_{\text{max}} \in [0,1]$, an \emph{uncertain directly-follows slice} is a function $\overline{DFG_{\text{U}}} \colon L \to \mathcal{U}_G$ where $\overline{DFG_{\text{U}}}(L, act_{\text{min}}, act_{\text{max}}, rel_{\text{min}}, rel_{\text{max}}) = (V, E)$ with:
	\begin{itemize}
		\item $V = \{a \in V' \mid act_{\text{min}} \leq \frac{\sum_{\sigma \in L} \#_\text{min}(\sigma, a)}{\sum_{\sigma \in L} \#_\text{max}(\sigma, a)} \leq act_{\text{max}} \}$
		\item $E = \{(a, b) \in E' \mid rel_{\text{min}} \leq \frac{\sum_{\sigma \in L} \rightsquigarrow_\text{min}(\sigma, a, b)}{\sum_{\sigma \in L} \rightsquigarrow_\text{max}(\sigma, a, b)} \leq rel_{\text{max}} \}$
	\end{itemize}
\end{definition}

A  UDFG slice is an unweighted directed graph which represents a filtering performed over vertices and edges of the UDFG. This graph can then be processed by the $\text{IM}_D$.

\begin{definition}[Uncertain Inductive Miner--directly-follows ($\text{UIM}_{D}$)]
	Given an uncertain event log $L \in \mathcal{P}(\mathcal{T}_U)$ and $act_{\text{min}}, act_{\text{max}}, rel_{\text{min}}, rel_{\text{max}} \in [0,1]$, the \emph{Uncertain Inductive Miner--directly-follows} ($\text{UIM}_{D}$) returns the process tree obtained by $\text{IM}_{D}$ over an uncertain DFG slice: $\text{IM}_{D}(\overline{DFG_{\text{U}}}(L, act_{\text{min}}, \allowbreak act_{\text{max}}, rel_{\text{min}}, rel_{\text{max}}))$.
\end{definition}

The filtering parameters $act_{\textit{min}}$, $act_{\textit{max}}$, $rel_{\textit{min}}$, $rel_{\textit{max}}$ allow to isolate the desired type of behavior of the process. In fact, $act_{\textit{min}} = rel_{\textit{min}} = 0$ and $act_{\textit{max}} = rel_{\textit{max}} = 1$ retain all possible behavior of the process, which is then represented in the model: both the behavior deriving from the process itself and the behavior deriving from the uncertain traces. Higher values of $act_{\textit{min}}$ and $rel_{\textit{min}}$ allow to filter out uncertain behavior, and to retain only the parts of the process observed in certain events. Vice versa, lowering $act_{\textit{min}}$ and $rel_{\textit{min}}$ allows to observe only the uncertain part of an event log.

\section{Experiments}\label{sec:exp}
The approach described here has been implemented using the Python process mining framework PM4Py~\cite{aless2019process}. The models obtained through the Uncertain Inductive Miner--directly-follows cannot be evaluated with commonly used metrics in process mining, since metrics in use are not applicable on uncertain event data; nor other approaches for performing discovery over uncertain data exist. This preliminary evaluation of the algorithm will, therefore, not be based on measurements; it will show the effect of the $\textit{UIM}_{D}$ with different settings on an uncertain event log.

Let us introduce a simplified notation for uncertain event logs. In a trace, we represent an uncertain event with multiple possible activity labels by listing the labels between curly braces. When two events have overlapping timestamps, we represent their activity labels between square brackets, and we represent the indeterminate events by overlining them. For example, the trace $\langle \overline{a}, \{b, c\}, [d, e] \rangle$ is a trace containing 4 events, of which the first is an indeterminate event with label $a$, the second is an uncertain event that can have either $b$ or $c$ as activity label, and the last two events have a range as timestamp (and the two ranges overlap). The simplified representation of the trace in Table~\ref{table:uncertaintrace} is $\langle [\{a, c\}, \{a, d\}], \overline{\{a, b\}}, [\{a, b\}, \{b, c\}], b \rangle$. Let us observe the effect of the $\textit{UIM}_{D}$ on the following test log:

$\langle a, b, e, f, g, h \rangle^{80}, \langle a, [\{b, c\}, e], \overline{f}, g, h, i \rangle^{15}, \langle a, [\{b, c, d\}, e], \overline{f}, g, h, j \rangle^{5}$.

\begin{figure}
	\centering
	\includegraphics[width=1\textwidth, keepaspectratio]{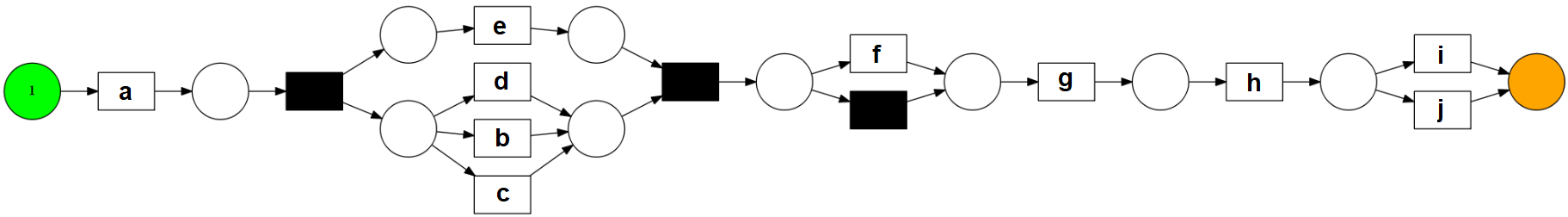}
	\caption{$\textit{UIM}_{D}$ on the test log with $act_{\textit{min}} = 0$, $act_{\textit{max}} = 1$, $rel_{\textit{min}} = 0$, $rel_{\textit{max}} = 1$.}
	\label{fig:net0-1-0-1}
\end{figure}

\begin{figure}
	\centering
	\includegraphics[width=1\textwidth, keepaspectratio]{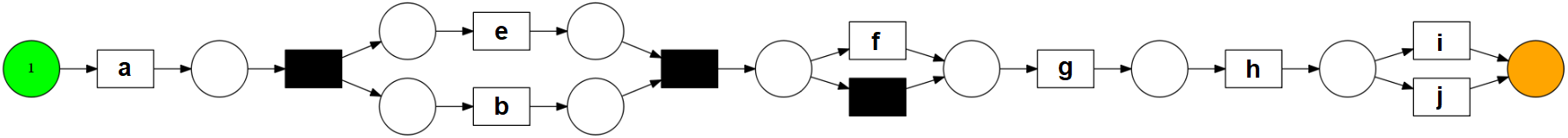}
	\caption{$\textit{UIM}_{D}$ on the test log with $act_{\textit{min}} = 0.6$, $act_{\textit{max}} = 1$, $rel_{\textit{min}} = 0$, $rel_{\textit{max}} = 1$.}
	\label{fig:net0.6-1-0-1}
\end{figure}

\begin{figure}
	\centering
	\includegraphics[width=0.7\textwidth, keepaspectratio]{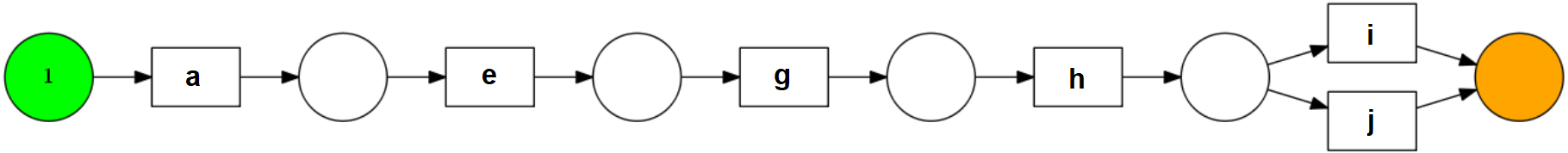}
	\caption{$\textit{UIM}_{D}$ on the test log with $act_{\textit{min}} = 0.9$, $act_{\textit{max}} = 1$, $rel_{\textit{min}} = 0$, $rel_{\textit{max}} = 1$.}
	\label{fig:net0.9-1-0-1}
\end{figure}

\begin{figure}
	\centering
	\includegraphics[width=1\textwidth, keepaspectratio]{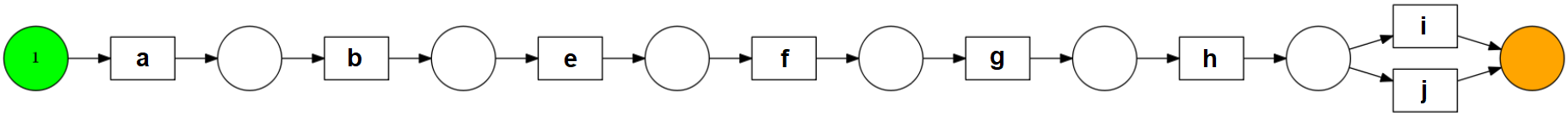}
	\caption{$\textit{UIM}_{D}$ on the test log with $act_{\textit{min}} = 0$, $act_{\textit{max}} = 1$, $rel_{\textit{min}} = 0.7$, $rel_{\textit{max}} = 1$.}
	\label{fig:net0-1-0.7-1}
\end{figure}

\begin{figure}
	\centering
	\includegraphics[width=0.5\textwidth, keepaspectratio]{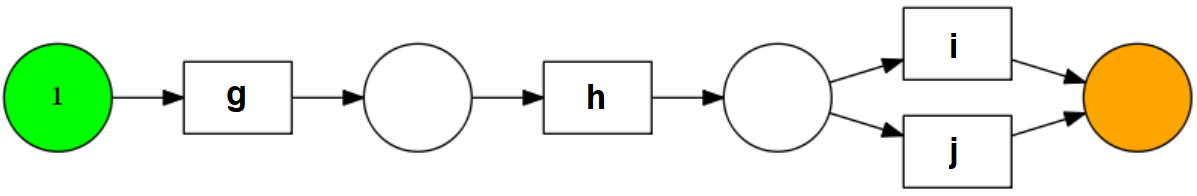}
	\caption{$\textit{UIM}_{D}$ on the test log with $act_{\textit{min}} = 0$, $act_{\textit{max}} = 1$, $rel_{\textit{min}} = 0.9$, $rel_{\textit{max}} = 1$.}
	\label{fig:net0-1-0.9-1}
\end{figure}

\begin{figure}
	\centering
	\includegraphics[width=1\textwidth, keepaspectratio]{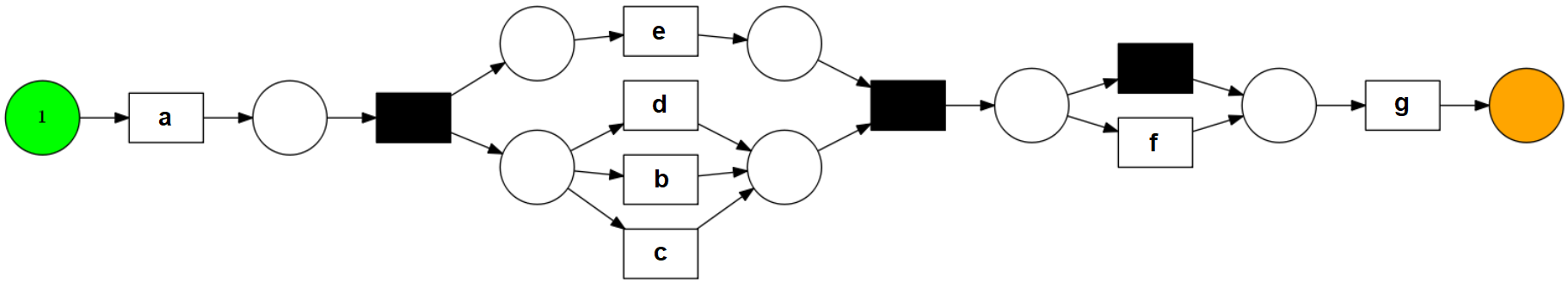}
	\caption{$\textit{UIM}_{D}$ on the test log with $act_{\textit{min}} = 0$, $act_{\textit{max}} = 1$, $rel_{\textit{min}} = 0$, $rel_{\textit{max}} = 0.8$.}
	\label{fig:net0-1-0-0.8}
\end{figure}

In Figure~\ref{fig:net0-1-0-1}, we can see the model obtained without any filtering: it represents all the possible behavior in the uncertain log. The models in Figures~\ref{fig:net0.6-1-0-1} and~\ref{fig:net0.9-1-0-1} show the effect on filtering on the minimum number of times an activity appears in the log: in Figure~\ref{fig:net0.6-1-0-1} activities $c$ and $d$ are filtered out, while the model in Figure~\ref{fig:net0.9-1-0-1} only retains the activities which never appear in an uncertain event (i.e., the activities for which $\#_\textit{min}$ is at least 90\% of $\#_\textit{max}$).

Filtering on $rel_{\textit{min}}$ has a similar effect, although it retains the most certain relationships, rather than activities, as shown in Figure~\ref{fig:net0-1-0.7-1}. An even more aggressive filtering of $rel_{\textit{min}}$, as shown in Figure~\ref{fig:net0-1-0.9-1}, allows to represent only the parts of the process which are never subjected to uncertainty by being in a directly-follows relationship that has a low $\rightsquigarrow_\textit{min}$ value.

The $\textit{UIM}_{D}$ allows also to do the opposite: hide certain behavior and highlight the uncertain behavior. Figure~\ref{fig:net0-1-0-0.8} shows a model that only displays the behavior which is part of uncertain attributes, while activities $h$, $i$ and $j$ -- which are never part of uncertain behavior -- have not been represented. Notice that $g$ is represented even though it always appeared as a certain event; this is due to the fact that the filtering is based on relationships, and $g$ is in a directly-follows relationship with the indeterminate event $f$.

\section{Conclusion}\label{sec:conc}
In this explorative work, we present the foundations for performing process discovery over uncertain event data. We present a method that is effective in representing a process containing uncertainty by exploiting the information into an uncertain event log to synthesize an uncertain model. The UDFG is a formal description of uncertainty, rather than a method to eliminate uncertainty to observe the underlying process. This allows to study uncertainty in isolation, possibly allowing us to determine which effects it has on the process in terms of behavior, as well as what are the causes of its appearance. We also present a method to filter the UDFG, obtaining a graph that represents a specific perspective of the uncertainty in the process; this can be then transformed in a model that is able to express concurrency using the $\textit{UIM}_{D}$ algorithm.

This approach has a number of limitations that will need to be addressed in future work. An important research direction is the formal definition of metrics and measures over uncertain event logs and process models, in order to allow for a quantitative evaluation of the quality of this discovery algorithm, as well as other process mining methods over uncertain logs. Another line of research can be the extension to the weakly uncertain event data (i.e., including probabilities) and the extension to event logs also containing uncertainty related to case IDs.

\bibliographystyle{splncs04}
\bibliography{discovering-process-models-from-uncertain-event-data}

\end{document}